\documentclass[reprint,superscriptaddress,
 amsmath,amssymb,
 aps,
]{revtex4-1}

\usepackage{graphicx}
\usepackage{dcolumn}
\usepackage{bm}

\begin{document}

\preprint{APS/123-QED}

\title{Multielectron effects in strong-field ionization of oriented OCS}

\author{Mahmoud Abu-samha}
\affiliation{College of Engineering and Technology, American University of the Middle East, Kuwait.}
\email{Mahmoud.Abusamha@aum.edu.kw}
\author{Lars Bojer Madsen}
\affiliation{Department of Physics and Astronomy, Aarhus University,  8000 Aarhus C, Denmark}

\date{\today}

\begin{abstract}
We present theoretical calculations of orientation-dependent total ionization yields from the highest occupied molecular orbitals of the oriented OCS molecule by solving the time-dependent Schr{\"o}dinger equation in three dimensions. The calculations were performed within the single-active-electron approximation including multielectron polarization. The multielectron polarization term was represented by an induced dipole term which contains the polarizability of the OCS$^+$ cation parallel to the laser polarization. Upon accounting for multielectron polarization, the calculated total ionization yields and their orientation dependence are in good agreement with experimental data.  
\end{abstract}

\maketitle


\section{\label{sec1}Introduction}
Recent advances in strong-field physics, allow identification of multielectron effects  in strong-field ionization experiments~\cite{Holmegaard2010,PhysRevLett.111.163001,Majety_2015b,PhysRevA.96.053421} and in  high-order harmonic generation experiments~\cite{Smirnova2009,Ferre2015}.
In parallel to experimental advances, theoretical models are currently being developed to account for multielectron polarization (MEP), see for example Ref.~\cite{PhysRevA.98.023406,PhysRevA.101.013433}.
All-electrons methods such as time-dependent Hartree-Fock theory (TDHF) and density functional theory (TDDFT)~\cite{PhysRevLett.52.997} have been used to model MEP in strong-field ionization of small atoms and molecules~\cite{PhysRevLett.111.163001,PhysRevA.98.043425}. These methods provide essential information on the contributions of different orbitals to the ionization process, and on the MEP effects caused by the interaction of the outgoing electron with the electron density of the parent ion. 

\begin{figure}
\includegraphics[width=0.5\textwidth]{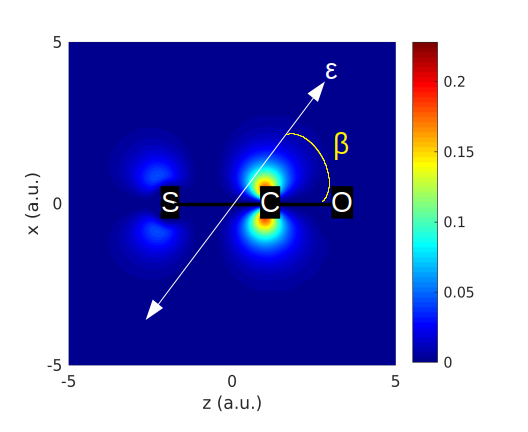}
\caption{\label{ocs_homo} A cross section of the electron density for the HOMO of the OCS molecule in the molecular-frame $xz$ plane. The HOMO in the $yz$ plane can be obtained by a $90^\circ$ rotation around the molecular axis. The OCS molecule is fixed at orientation angle, $\beta$, relative to linear laser polarization, $\varepsilon$. }
\end{figure}

Strong-field ionization of the OCS molecule has received considerable attention in the past decade and can be considered as an important test case for the elucidation of MEP effects~\cite{HansenJPhysB2011,Johansen_2016,PhysRevA.98.043425,PhysRevA.87.013406,PhysRevA.89.013405,PhysRevA.98.043425,Yu_2017}. For the oriented OCS molecule, while the total ionization yield (TIY) is dominated by ionization from the degenerate highest occupied molecular orbitals (HOMOs), see Fig.~\ref{ocs_homo}, the ionization process is strongly affected by MEP~\cite{PhysRevA.98.043425}. 
For the OCS molecule, a series of experimental measurements were conducted for the investigation of the TIYs and their orientation dependence~\cite{PhysRevA.89.013405,PhysRevA.98.043425,HansenJPhysB2011,Yu_2017,Johansen_2016}. In Ref.~\cite{HansenJPhysB2011}, the experiments were conducted using linearly polarized laser pulses with 30-fs duration at 800-nm wavelength and intensities of 1.5$\times10^{14}$~W/cm$^2$ and 1.8$\times10^{14}$~W/cm$^2$. In these experiments, the TIY was measured as a function of the orientation angle $\beta$, see Fig.~\ref{ocs_homo} which illustrates the orientation of the OCS molecule relative to the linear laser polarization. The minimum TIY was obtained with the laser polarization parallel to the molecular axis at $\beta=0^\circ$, whereas the maximum TIY was obtained with the laser polarization perpendicular to the molecular axis at $\beta=90^\circ$. 
While the TIY is dominated by ionization from the HOMO of the OCS molecule, the orientation dependence of the TIYs presents a challenge to ionization models. For example, the molecular Ammosov-Delone-Krainov model~\cite{PhysRevA.66.033402} and the strong-field approximation (SFA)~\cite{Keldysh,Faisal,Reiss}, predict that the orientation dependence of the TIY should follow the orbital structure. Hence in these models, the TIY should peak at $\beta$ close to 35$^\circ$, see electron density plot of the HOMO in Fig.~\ref{ocs_homo}. Even the weak-field asymptotic theory of tunneling, which includes dipole effects~\cite{Tolstikhin2011}, does not capture the experimental result and predicts a maximum at $\beta \simeq 30^\circ$~\cite{PhysRevA.87.013406}.
Likewise the Stark-corrected molecular tunneling theory~\cite{PhysRevA.82.053404,Holmegaard2010} does not reproduce these experimental results for the OCS molecule: the Stark-corrected molecular tunneling theory predicts the maximum ionization yield at $\beta=45^\circ$\cite{PhysRevA.83.023405}. The results of Ref.~\cite{HansenJPhysB2011} suggest that the angular behaviour of the TIYs in
linearly polarized fields could not be explained by ionization rates only. In fact, to map out the instantaneous ionization rates of the OCS molecule, circularly-polarized laser pulses had been used in previous studies~\cite{Holmegaard2010,PhysRevA.83.023405} because both re-scattering with the parent ion and the influence of excited molecular states is minimized for this choice of polarization.

From the previous discussion, it is evident that an accurate description of strong-field ionization of oriented OCS  by  linearly-polarized light requires a theory, which accounts for multielectron effects and propagation of the outgoing electron in the combined potential of the remaining cation and the laser pulse. In Ref.~\cite{PhysRevA.98.043425}, a new set of measurements were presented for OCS using 800-nm linearly-polarized laser pulses with 30-fs duration and peak intensities of 4.5$\times 10^{13}$~W/cm$^2$, 7$\times 10^{13}$~W/cm$^2$, and 2$\times 10^{14}$~W/cm$^2$. In addition to the experimental measurements, the authors presented calculations of TIYs within the framework of TDDFT. The computational results were in a good agreement with the experimental measurements and explained the failure of the molecular tunneling theory to describe the orientation dependence of the TIYs by a multichannel ionization mechanism, which was revealed by the localization of a hole on the sulfur atom following strong-field ionization in the parallel orientation (at $\beta=0^\circ$), a mechanism that can not be captured by ionization models such a single-electron tunneling theory or the SFA. Although the TDDFT method captures essential parts of the dynamics of strong-field ionization of oriented OCS, it can suffer from inherent problems such as the inability to distinguish between single and double ionization yields, and be challenged in the accuracy of results at low intensities due to electron leakage. In addition, it can be hard to obtain a good representation of contributions from electronically excited states in the multiphoton ionization process. This latter point is noteworthy in light of the recent results of Ref.~\cite{Yu_2017}, in which  strong-field ionization of oriented OCS  revealed a  contribution from resonant excitation in the photoelectron spectra and angular distributions at different laser intensities and wavelengths.
 
In view of the interests and challenges in the response of OCS to an intense laser pulse,  we find it interesting to revisit this problem theoretically with a new TDSE methodology, which can handle MEP within the single-active-electron (SAE) approximation~\cite{PhysRevA.101.013433}. This approach will be computationally much cheaper than any multielectron methodology and will allow a clear and physically transparent identification of the influence of the MEP term. Moreover,  
in cases where MEP is of central importance in the dynamics such as in OCS, 
the performance of the present SAE approximation is expected to be of the same quality if not superior to all-electron methods for the following reason: in all-electron methods, which inevitably introduce approximations for the electron-electron correlation, the accuracy of the polarizability of the ion is determined by the level of theory implemented in the method. By contrast, in the SAE model, the polarizability of the ion is an input parameter which can be chosen as the most accurate experimental or theoretical value available. Here, we calculate TIYs from the HOMO of oriented OCS as a function of the orientation angle, using the TDSE within the SAE approximation and including MEP corrections. The calculations were performed for OCS in linearly-polarized laser pulses which contain 5 optical cycles at 800-nm wavelength and with peak intensities of 8$\times10^{12}$~W/cm$^2$, 2.2$\times10^{13}$~W/cm$^2$, and 4.5$\times10^{13}$~W/cm$^2$. The MEP was represented using an induced-dipole term based on the polarizability of the OCS$^+$ ion. When the MEP term is included, the TDSE calculations predict the maximum TIY at the orientation $\beta=90^o$ in agreement with the experimental measurements of Refs.~\cite{HansenJPhysB2011,PhysRevA.98.043425}.

The theoretical and computational models are presented in Sec.~\ref{compdet}, followed by results and discussion in Sec.~\ref{res} and conclusions in Sec.~\ref{conc}. Atomic units are used throughout unless otherwise stated.

\section{Theoretical and Computational Models}
\label{compdet}
This section is divided into three parts. In the first part, we describe the procedures for generating the SAE potential and the wavefunction for the HOMO of OCS. In the second part, we give a brief description of our TDSE method. In the final part,  we describe how to include MEP in the TDSE method within the SAE approximation.

\subsection{Single-active-electron potential and the HOMO of OCS}
\label{sec:sae}
The SAE potential describing OCS was obtained from quantum chemistry calculations following the procedure described in Ref.~\cite{PhysRevA.81.033416}. The molecule is placed along the molecular-frame $z$-axis such that the center-of-mass coincides with the origin and the oxygen (sulfur) atom points in the positive (negative) $z$ direction. The SAE potential of OCS was expanded in partial waves as $V(\vec{r})=\sum_{l,m=0}^{l_{max}}V_{l0}(r)Y_{l0}(\theta,\phi)$ where $m=0$ since the molecule is linear and the potential is invariant under rotations around the molecular axis. The expansion of the molecular potential was truncated at $l_{max}=20$. Based on our SAE potential for the OCS molecule, the HOMO of OCS is a $\pi$ orbital with energy -0.35~a.u., which is slightly above a reference orbital energy of -0.42 a.u. as obtained from quantum chemistry calculations using standard quantum chemistry software (GAMESS~\cite{gamess}).

We applied the split-operator spectral method~\cite{PhysRevA.38.6000} to obtain the wavefunction corresponding to the HOMO of OCS. We started the time propagation from an inital guess wavefunction, denoted by $\psi(\vec{r},t=0)$ and used the SAE potential for OCS. For the inital state, we need a state that includes the $\pi$ orbital symmetry and we chose the hydrogenic 2p$_x$ state for convenience. We performed field-free propagation for a time duration of 1000 a.u. and saved the wavepacket every 1 a.u. of time. From the time-dependent wavefunction obtained in this manner, $\psi(\vec{r},t)$, an autocorrelation function $\mathcal{P}(t)= \langle\psi(\vec{r},t=0)|\psi(\vec{r},t) \rangle$ was calculated, and a bound state spectrum $\mathcal{P}(E)$ was obtained as 
\begin{equation}
\mathcal{P}(E) =     \frac{1}{T}\int_{0}^{T}dt~w(t)~\exp(iEt)~\mathcal{P}(t),
\end{equation}
where $w(t)$ is the Hanning window function~\cite{PhysRevA.38.6000}. Once the HOMO energy, $E_\text{HOMO}$, is well resolved in the $\mathcal{P}(E)$ spectrum, the corresponding wavefunction can be constructed and normalized as follows
\begin{equation}
\label{wfc_homo}
\psi_\text{HOMO}(\vec{r}) =     \frac{1}{T}\int_{0}^{T}dt~w(t)~\exp(iE_\text{HOMO}t )\Psi(\vec{r},t).
\end{equation}
We used the wavefunction of Eq.~(\ref{wfc_homo}) as the initial state for the solution of the TDSE.

\subsection{Summary of the TDSE calculations}
The TDSE methodology was thoroughly discussed elsewhere~\cite{Kjeldsen2007a}. Here, we give a brief overview of the method. The TDSE is solved for the active (HOMO) electron within the SAE approximation. The time-dependent wavefunction, $\psi(\vec{r},t)$, is expanded in spherical harmonics $Y_{lm}(\Omega)$ for the angular degrees of freedom and a radial grid for the time-dependent reduced radial wave functions, $f_{lm}(r,t)$, i.e.,  
\begin{equation}
    \label{eq:wf}
\psi(\vec{r},t)=\sum_{lm} \frac{f_{lm}(r,t)}{r} Y_{lm}(\Omega). 
\end{equation}
The TDSE is propagated in the length gauge (LG) with a combined split-operator~\cite{PhysRevA.38.6000} Crank-Nicolson method. In the LG, the interaction of the laser field with the active electron is given by $V_{\text{Ext}}(\vec{r},t)=\vec{E}(t) \cdot \vec{r}$. The electric field $\vec{E}(t)$, linearly polarized along the lab-frame $z-axis$, is defined as 

\begin{equation}
    \vec{E}(t) = -\partial_t \vec{A}(t) \hat{z}= -\partial_t \left(\frac{E_0}{\omega}\sin^2(\pi t/\tau)\cos(\omega t+\phi) \right)\hat{z},
    \label{E_field}
\end{equation}
where $\vec{A}(t)$ is the vector potential, $E_0$ is the field amplitude, $\omega$ the frequency, and $\phi$ the carrier-envelope phase (CEP) for a laser pulse with duration $\tau$.  The construction of the laser electric field from the above vector potential, ensures that there are no unphysical DC components in the field~\cite{Madsen2002}.

The TDSE calculations were performed at a laser frequency of $\omega=0.057$~a.u. corresponding to a wavelength of 800~nm, a CEP value of $\phi=-\pi/2$, and the pulses contained 5 cycles with field amplitudes ($E_0=0.015$, $0.025$, and $0.036$~a.u.) corresponding to peak intensities of $8\times10^{12}$, $2.2\times10^{12}$ and $4.5\times10^{13}$ W/cm$^2$. We chose somewhat lower peak intensities than what is reported experimentally in order to achieve convergence of the TDSE calculations, in particular since our SAE potential underestimates the ionization potential of the HOMO for OCS, see Sec.~\ref{sec:sae}. In the TDSE calculations, the radial grid contained 8192 points and extended to 320~a.u. The size of the angular basis set was limited by setting $l_{max}=60$ in the partial wave expansion of the wavefunction, see Eq.~(\ref{eq:wf}). The calculations were performed at orientation angles in the range $\beta=0^\circ-90^\circ$  with a step of 15$^\circ$. The results have been checked for convergence by varying $l_{\max}$.

The TIYs 
were produced by projecting the wavepacket at the end of the laser pulse on Coulomb scattering states in the asymptotic region ($r>20$~a.u.), an approach that was validated in Ref.~\cite{Madsen2007} and recently applied for the molecular hydrogen ion~\cite{PhysRevA.94.023414,1742-6596-869-1-012009} and the polar CO molecule~\cite{PhysRevA.101.013433}.

\subsection{Including the MEP term in the TDSE calculations}
Theory accounting for the effect of MEP in strong-field ionization was developed in Refs.~\cite{PhysRevLett.95.073001,doi:10.1080/09500340601043413,PhysRevA.82.053404,Holmegaard2010,0953-4075-51-10-105601}. To include MEP in our TDSE method, we adopted the approach layed out in Ref.~\cite{0953-4075-51-10-105601}, see also Ref.~\cite{PhysRevA.101.013433}. For the polar OCS molecule, the potential describing the interaction of the active electron with the core including the MEP term and the time-dependent external field, is given asymptotically as~\cite{PhysRevLett.95.073001,PhysRevA.82.053404,PhysRevA.82.043413}
\begin{equation}
\label{saepot2}
V_\text{eff}(\vec{r},t)|_{r \rightarrow \infty} = \vec{r}\cdot\vec{E}(t) -\frac{Z}{r}  - \frac{(\vec{\mu}_p + \vec{\mu}_{ind})\cdot \vec{r} }{r^3} + \dots ,
\end{equation}
where $Z$=1 is the charge of the cation,  $\vec{\mu}_p$ and $\vec{\mu}_{ind}$ are the permanent and induced dipoles of the OCS$^+$ cation. The MEP term is defined as $- \vec{\mu}_{ind}\cdot \vec{r} /r^3 = -\alpha_{||} \vec{E}(t)\cdot\vec{r}/r^3 $ where $\alpha_{||}$ is the static polarizability of the OCS$^+$ ion parallel to the laser polarization axis. A cutoff radius is chosen close to the core at a radial distance
\begin{equation}
   r_{c}=\alpha_{||}^{1/3}
   \label{r_c}
\end{equation}
such that the MEP cancels the external field at $r\le r_c$~\cite{0953-4075-51-10-105601,PhysRevLett.95.073001,doi:10.1080/09500340601043413}.

\begin{figure*}
\includegraphics[width=1.0\textwidth]{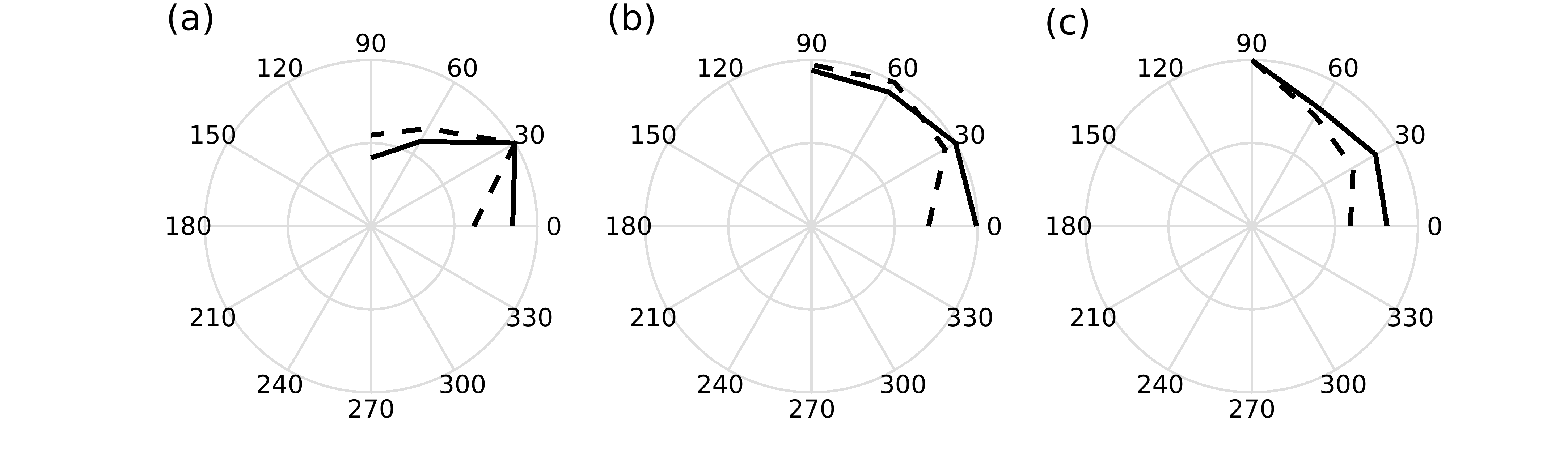}
\caption{\label{ocs_tiy_mepeff} Orientation-dependent TIYs from the HOMOs of the OCS molecule probed by linearly polarized laser pulses containing 5 optical cycles at 800 nm with different models of MEP. In (a), the isotropic polarizability of the OCS$^+$ ion is used at peak laser intensities of  $8\times 10^{12}$ W/cm$^2$ (dashed curve) and $2.2\times 10^{13}$ W/cm$^2$ (solid curve). In (b), the external field is turned off within $r_c = \alpha_{||}^{1/3}$ while neglecting the induced dipole term ($-\alpha_{||} \vec{E}(t)\cdot\vec{r}/r^3$) at peak laser intensities  of  $2.2\times 10^{13}$ (dashed curve) and $4.5\times 10^{13}$ W/cm$^2$ (solid curve). In (c), the full MEP effect is accounted for  based on the parallel polarizability of the OCS$^+$ ion ($\alpha_{||}$) at peak laser intensities of $2.2\times 10^{13}$ W/cm$^2$ (dashed curve) and $4.5\times 10^{13}$ W/cm$^2$ (solid curve). The maxima in the TIYs (normalized to unity in the plots for a better comparison) at the low and high intensities, respectively are as follows: (a) $2.58 \times 10^{-6}$ and $2.17 \times 10^{-4}$ (b) $1.15 \times 10^{-3}$ and $1.83 \times 10^{-2}$, (c) $4.52 \times 10^{-4}$ and $7.77 \times 10^{-3}$.}
\end{figure*}
Implementing the MEP term in the TDSE method is straightforward in the LG. In our approach, upon including MEP, the interaction term is expressed in the LG at each radial point $r$ as~\cite{PhysRevA.101.013433}
\begin{equation}
\label{Eq2}
    V_{LG}^{Ext}(r,t)= \begin{cases}
  \left(1-\frac{\alpha_{||}}{r^3}\right)\vec{E}(t)\cdot\vec{r};~r > r_c \\ 
  0;~r \le r_c \\
  \end{cases}
  \end{equation}
The interaction term is zero at $r\le r_c$ because the external field is counteracted by the MEP~\cite{PhysRevLett.95.073001,doi:10.1080/09500340601043413}. Note that the permanent dipole term of Eq.~\eqref{saepot2} is already part of the field-free SAE potential of OCS.

The MEP term was implemented in the TDSE calculations in the LG as discussed in detail in Ref.~\cite{PhysRevA.101.013433}. The polarizability of the OCS$^+$ ion was taken from the NIST computational chemistry database~\cite{cccbdb}, as computed using the BLYP functional of density functional theory and the cc-pVDZ basis set in the geometry of the ion. 
At this level of theory, the OCS$^+$ ion has a static polarizability with the following nonzero components: $\alpha_{xx}=12.56,~\alpha_{yy}=12.96,~\alpha_{zz}=40.72$~a.u., whereas the nonzero component of the dynamic polarizability (computed at 800 nm) are as follows: $\alpha_{xx}=13.16,~\alpha_{yy}=12.57,~\alpha_{zz}=41.78$~a.u. The small differences between the static and dynamic polarizability components suggest that the static polarizability is suitable for the present calculations. For the static polarizability, the spherical component is 21.94~a.u. whereas the anisotropic component is 28.16~a.u. We have checked that calculating the polarizability of the OCS$^+$ ion in the geometry of the neutral molecule results in very similar values for the above polarizability components. The polarizability component parallel to the laser polarization axis is defined as
\begin{equation}
\label{alpha_beta}
\alpha_{||}=\cos^2(\beta)\alpha_{zz} +\sin^2(\beta)\alpha_{xx},
\end{equation}
where $\beta$ is the orientation angle, defined in Fig.~\ref{ocs_homo}. Accordingly, the value of the polarizability parallel to the laser polarization is $\alpha_{||}=\alpha_{zz}$=40.72~a.u. at $\beta=0^\circ$ ($\beta=180^\circ$) when the laser polarization is parallel to the molecular axis, whereas $\alpha_{||}=\alpha_{xx}$=12.56~a.u. at $\beta=90^\circ$ when the laser polarization is perpendicular to the molecular axis. The large polarizability anisotropy between the orientation angles $\beta=0^\circ$ and $90^\circ$ is essential in order to correctly capture the MEP effect in strong-field ionization of OCS.

\section{Results and Discussion}
\label{res}
%

In this section, we will discuss the effects of MEP on the TIYs for the oriented OCS molecule. The TDSE calculations were performed within the SAE approximation including the effect of MEP. The laser pulses, linearly polarized, contained five optical cycles at a wavelength of 800 nm. The TIYs were calculated at the end of the laser pulses for the degenerate HOMOs and the overall TIY is obtained as the sum of the contributions from the two HOMOs. Notice that at $\beta=0^\circ$, the external laser field runs through the molecular axis, and due to cylindrical symmetry, both HOMOs produce the same TIYs at this orientation.  Although the contribution to the overall TIY from the HOMO in the molecular $yz$ plane is small, it can affect the position of the orientation angle of maximum ionization yield and that is why its contribution should be considered. 
\begin{figure*}
\includegraphics[width=1.0\textwidth]{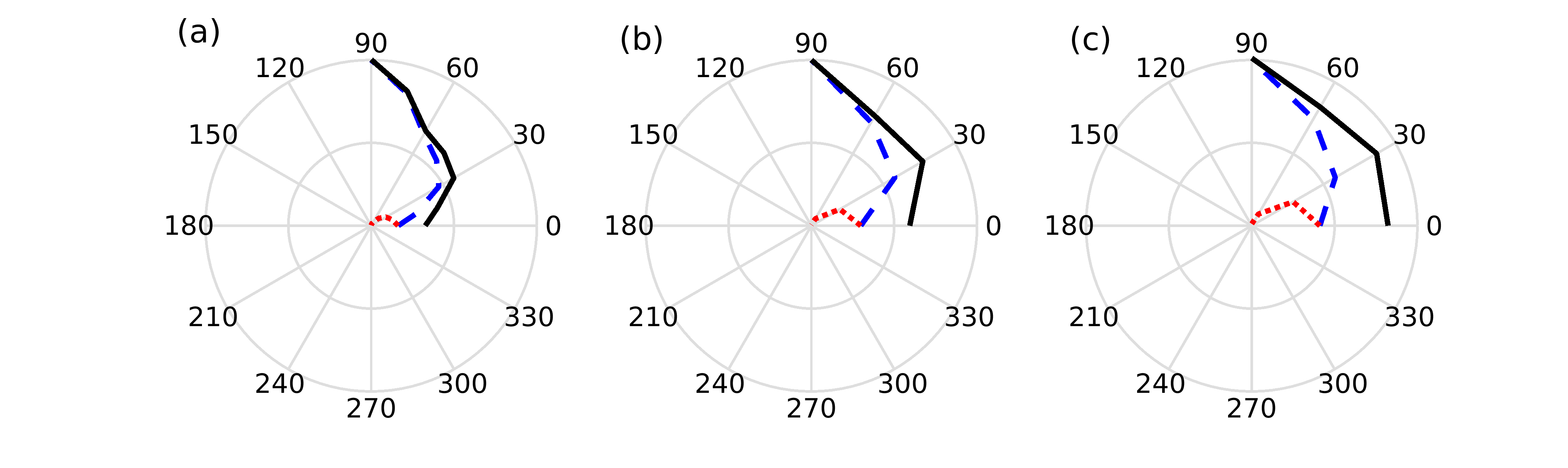}
\caption{\label{ocs_tiy_inteff} Orientation-dependent TIYs from the degenerate HOMOs of the OCS molecule probed by linearly polarized laser pulses containing 5 optical cycles at 800 nm with peak intensities of (a) $8\times 10^{12}$ W/cm$^2$,(b) $2.2\times 10^{13}$ W/cm$^2$, and (c) $4.5\times 10^{13}$ W/cm$^2$. The blue dashed (red dotted) curves denote the contribution from the HOMO in the $xz$-plane ($yz-$plane), the black curves denote the summed-up TIYs. The maxima in the TIYs (normalized to unity in the plots for a better comparison) are (a) $5.53 \times 10^{-6}$, (b) $4.52 \times 10^{-4}$, and (c) $7.77 \times 10^{-3}$.}
\end{figure*}

As a first attempt to include MEP effects, the TDSE calculations were conducted using the isotropic polarizability of the OCS$^+$ ion, $\alpha=21.95$ a.u.~\cite{cccbdb}, and its corresponding cutoff radius $r_c=2.8$ a.u., see Eq.~(\ref{r_c}). This value of the polarizability and the corresponding $r_c$ were then applied at all orientation angles. The TDSE calculations were conducted at laser intensities of $8\times 10^{12}$ W/cm$^2$ and $2.2\times 10^{13}$ W/cm$^2$. The summed up TIYs for the degenerate HOMOs are plotted at different orientation angles in Fig.~\ref{ocs_tiy_mepeff}~(a). From Fig.~\ref{ocs_tiy_mepeff}~(a), one can see that the maximum ionization yield is obtained at $\beta=30^\circ$ at both laser intensities. Clearly, using the isotropic polarizability of the OCS$^+$ ion results in a maximum ionization yield at too small orientation angle in comparison with the experimental results~\cite{HansenJPhysB2011,PhysRevA.98.043425} where the maximum ionization yield is expected at $\beta=90^\circ$. This disagreement indicates that accounting for MEP effects using an isotropic polarizability model is not satisfactory. The results show that the polarizability anisotropy should be considered. Notice that polarizability anisotropy means that the parallel $\beta$-dependent polarizability value, $\alpha_{||}$ of Eq.~\eqref{alpha_beta}, and the corresponding cutoff radius, $r_c$ of Eq.~\eqref{r_c}, should be considered. 

The TDSE calculations were then conducted using the parallel polarizability of the OCS$^+$ ion, which accounts for polarizability anisotropy. At first, we considered switching off the external field within a radial cutoff distance $r_c=\alpha_{||}^{1/3}$ determined based on the polarizability of the OCS$^+$ along the laser polarization, without including an induced dipole term in the MEP model, i.e., the term $-\alpha_{||} \vec{E}(t)\cdot\vec{r}/r^3$ in Eq.~(\ref{Eq2}) was omitted. The TDSE calculations were conducted at laser intensities of $2.2\times 10^{13}$ W/cm$^2$ and
$4.5\times 10^{13}$ W/cm$^2$. The summed-up TIYs for the degenerate HOMOs are presented in Fig.~\ref{ocs_tiy_mepeff}~(b). From  Fig.~\ref{ocs_tiy_mepeff}~(b) at both intensities, although the ionization yield does not change significantly with the orientation angle, in particular at the higher intensity, we do not observe a maximum yield at $\beta=90^\circ$. The orientation angle of maximum TIY depends on laser intensity: the maximum yield is observed at $\beta=60^\circ$ at $2.2\times 10^{13}$ W/cm$^2$ and at $\beta=30^\circ$ at $4.5\times 10^{13}$ W/cm$^2$. Notice that in the reported experimental measurements~\cite{HansenJPhysB2011,PhysRevA.98.043425}, the orientation angle of maximum ionization yield ($\beta=90^\circ$) does not depend on laser intensity. A careful  analysis of the TIYs and their orientation dependence [see Fig.~\ref{ocs_tiy_mepeff}~(b)] suggests that since the induced dipole term, $-\alpha_{||} \vec{E}(t)\cdot\vec{r}/r^3$ in  Eq.~(\ref{Eq2}), was neglected in the TDSE calculations, the TIYs calculated at small orientation angles must be significantly overestimated, in particular at $\beta=0^\circ$. This, in turn, resulted in overestimating the contribution to the TIYs from the HOMO in the $yz-$plane. To investigate this aspect further, our next step is to take the full MEP effect into consideration.  

We account for the full MEP effect in the TDSE calculations; where both the short range cutoff radius of Eq.~\eqref{r_c} and the long-range induced-dipole term, $-\alpha_{||} \vec{E}(t)\cdot\vec{r}/r^3$ in  Eq.~(\ref{Eq2}), are included based on the $\beta$-dependent parallel polarizability of the OCS$^+$ cation, see Eq.~\eqref{alpha_beta}. 
The summed-up TIYs for the degenerate HOMOs are shown in Fig.~\ref{ocs_tiy_mepeff}~(c). As can be seen, the TIY is maximum at $\beta=90^\circ$, at both laser intensities, in excellent agreement with the experimental findings~\cite{HansenJPhysB2011}.


In Fig.~\ref{ocs_tiy_inteff}, we compare the orientation-dependent TIYs from the degenerate HOMOs  of the OCS molecule at three different intensities of $8\times 10^{12}$ W/cm$^2$, $2.2\times 10^{13}$ W/cm$^2$, and $4.5\times 10^{13}$ W/cm$^2$. In Fig.~\ref{ocs_tiy_inteff}, in addition to the summed-up TIYs, we show the separate contributions from the degenerate HOMOs: the ionization yields corresponding to the HOMO in the molecular $xz$ plane are represented by blue dashed curves, while the yields corresponding to the HOMO in the molecular $yz$ plane are represented by red dotted curves. The TDSE calculations were performed using the parallel polarizability of the OCS$^+$ ion of Eq.~\eqref{alpha_beta}. From Fig.~\ref{ocs_tiy_inteff}, one can see that at all intensities the HOMO in the molecular $xz$ plane has its maximum yield at $\beta=90^\circ$ with the external field perpendicular to the molecular axis, whereas the HOMO in the molecular $yz$ plane has its maximum yield at $\beta=0^\circ$ with the external field parallel to the molecular axis. Moreover, the maximum overall TIY is obtained at $\beta=90^\circ$ at all intensities. While the intensity has no effect on the orientation angle of maximum TIY, it definitely has an effect on the ratio ($\mathcal{R}$) between the ionization yields at $\beta=0^\circ$ and $\beta=90^\circ$, where $\mathcal{R}$ is defined as $\mathcal{R}=\frac{\text{TIY}(\beta=0^\circ)}{\text{TIY}(\beta=90^\circ)}$~\cite{HansenJPhysB2011}. Based on the results in Fig.~\ref{ocs_tiy_inteff}, the ratio of ionization yields has values of $\mathcal{R}\approx0.3$ at $8\times 10^{12}$ W/cm$^2$, $\mathcal{R}\approx0.6$ at $2.2\times 10^{13}$ W/cm$^2$ and $\mathcal{R}\approx0.8$ at $4.5\times 10^{13}$ W/cm$^2$. This trend in $\mathcal{R}$ and its dependence on laser intensity is consistent with both the TDDFT calculations and the experimental measurements on OCS reported in Ref.~\cite{PhysRevA.98.043425}, in which the ratio between the ionization yields at $\beta=0^\circ$ and $\beta=90^\circ$ degrees increases with increasing intensities. Our estimates of $\mathcal{R}$ at the higher intensities are comparable to the experimental value of 0.65~\cite{HansenJPhysB2011} determined at a laser intensity of $1.5\times 10^{14}$ W/cm$^2$.

\section{Conclusion}
\label{conc}
For the oriented OCS molecule,  TIYs 
were obtained from TDSE calculations for the degenerate HOMOs within the SAE approximation and taking into account MEP. In the treatment of MEP, the induced dipole term was first calculated based on the isotropic polarizability of the OCS$^+$ ion, in which case the resultant ionization yields failed to reproduce the experimental measurements~\cite{HansenJPhysB2011,PhysRevA.98.043425}. However, when the induced dipole term was calculated based on the full polarizability of the OCS$^+$, in which the polarizability parallel to the laser polarization and the corresponding cutoff radius depends on the orientation angle, a good agreement was obtained between our calculated TIYs and the experimental measurements~\cite{HansenJPhysB2011,PhysRevA.98.043425}. The calculations were performed at several intensities and the agreement with the experiment is satisfactory at all considered intensities. These findings show that the SAE methodology can capture fundamental aspects of multielectron dynamics in strong-field ionization of the OCS molecule when suitably extended to account for the rearrangement of the remaining core-electrons as described by the induced dipole potential.


\begin{acknowledgments}
The numerical results presented in this work were obtained at the Centre for Scientific Computing (CSCAA), Aarhus. 
\end{acknowledgments}


\providecommand{\noopsort}[1]{}\providecommand{\singleletter}[1]{#1}%

\end{document}